\documentclass[letterpaper,prx,longbibliography,accepted=2024-07-15]{quantumarticle}
\pdfoutput=1
\usepackage[utf8]{inputenc}
\usepackage[english]{babel}
\usepackage[T1]{fontenc}
\usepackage{amsmath}
\usepackage{hyperref}

\usepackage{tikz}
\usepackage{lipsum}

\usepackage{graphicx}
\usepackage{latexsym}
\usepackage{amsthm}
\usepackage{amssymb}
\usepackage{epstopdf} 
\usepackage{mathtools} 
\usepackage{enumitem}
\usepackage{setspace}
\usepackage{dcolumn}
\usepackage{bm}
\usepackage{color}

\usepackage[numbers]{natbib}

\DeclareFontFamily{OT1}{pzc}{}
\DeclareFontShape{OT1}{pzc}{m}{it}{ <-> s*[1.3] pzcmi7t }{}
\DeclareMathAlphabet{\mathpzc}{OT1}{pzc}{m}{it}

\begin{document}

\title{Mixed-State Quantum Spin Liquids and Dynamical Anyon Condensations
in Kitaev Lindbladians}

\author{Kyusung Hwang}
\affiliation{School of Physics, Korea Institute for Advanced Study, Seoul 02455, Korea}
\orcid{0000-0003-0910-5522}
\email{khwang@kias.re.kr}
\maketitle

\begin{abstract}
Quantum spin liquids and anyons, used to be subjects of condensed matter physics, now are realized in various platforms of qubits, offering unprecedented opportunities to investigate fundamental physics of many-body quantum entangled states. Qubits are inevitably exposed to environments’ effects such as decoherence and dissipation, which are believed to be detrimental to many-body entanglement.
Here, we argue that unlike the common belief decoherence and dissipation can give rise to novel topological phenomena in quantum spin liquids.
We study open quantum systems of the Kitaev spin liquid and the toric code via the Lindblad master equation approach. By using exact solutions and numerical approaches, we show the dynamical occurrence of anyon condensation by decoherence and dissipation, which results in a topological transition from the initial state spin liquid to the steady state spin liquid.
The mechanism of the anyon condensation transition by the Lindblad dynamics is elucidated.
We also provide an insight into the relationship between the Kitaev spin liquid and the toric code in the picture of anyon condensation.
Our work suggests open quantum systems to be a new venue for topological phenomena of quantum spin liquids and anyons.
\end{abstract}

\section{Introduction}

Quantum spin liquids are exotic phases of matter featured with emergent anyon quasiparticles and long range entanglement~\cite{Kitaev2006,Kitaev2003,WenBook,SachdevBook,Savary2016,Broholm2020,Knolle2019,Kasahara2018,Takagi2019,Motome2019,Trebst2022}. In addition to their importance in fundamental physics, anyons and quantum spin liquids have been a central topic of recent studies due to their potential applications in quantum memories and quantum computations~\cite{Preskill2002,Kitaev2003,Kitaev2006,Bombin2008,Bombin2010,Pachosbook,Terhal2015,Jian-WeiPan2017,Satzinger2021,Semeghini2021,Samajdar2020,Verresen2021,Williamson2022,Pollmann2022,Hsieh2022,Tantivasadakarn2023,Google2023,Verresen2022,Bravyi2022,Tantivasadakarn2022shortest,Kim2023,Iqbal2023}.
In the aspect of applications, it is essential to understand quantum devices under the unavoidable influences of environments such as decoherence and dissipation effects~\cite{Preskill2018}. Studies of such open quantum systems also provide a promising opportunity for exploring new phenomena fostered by many-body quantum entanglement and the environment effects~\cite{Zoller2008,Diehl2008natphys,Diehl2011natphys}. Nonetheless, most studies of quantum spin liquids have focussed on closed systems so far. Open quantum spin liquids remain largely unexplored due to the complexities of the problems and the absence of solvable systems.

In this work, we introduce open quantum spin liquids that are exactly solved in the steady state limit. As a primary example, we consider the Kitaev spin liquid (KSL)~\cite{Kitaev2006} coupled with a Markovian environment via the Lindblad master equation approach~\cite{Lindblad1976,Gorini1976,BreuerPetruccione,Albert2014,Albert2016,Lidar2020,Harrington2022}.
Under the Lindblad dynamics, the pure state of the KSL evolves to a steady state that shows vanishing spin-spin correlation, yet still preserving the zero-flux quantum number. The exact form of the steady state is given by the maximally mixed state in the zero-flux sector with equal weight. We call this state ``mixed-state Kitaev spin liquid''.

\begin{figure*}[tb]
\centering
\includegraphics[width=0.6\linewidth]{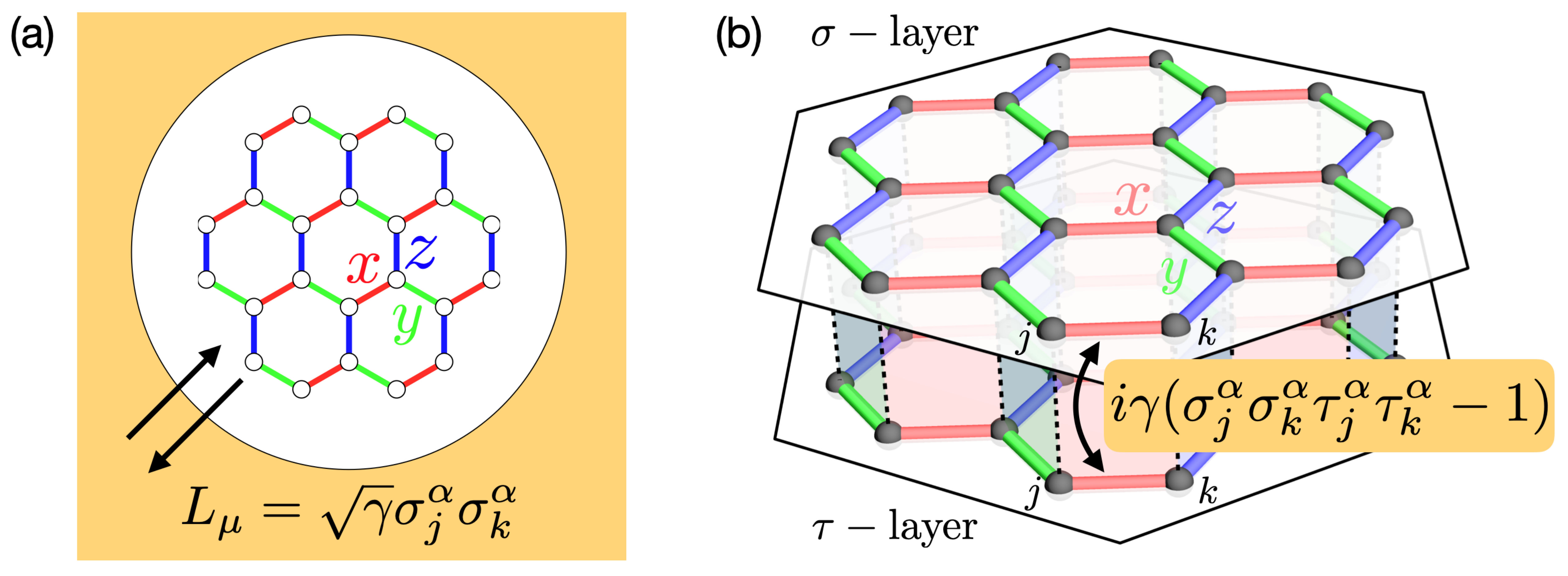}
\begin{tabular}{lcc}
\hline
& Original system & Doubled system
\\
\hline
State & $\hat{\rho}$ & $|\rho \rangle\!\rangle$
\\
Time evolution & $\frac{d}{d t} \hat{\rho} = \mathcal{L}(\hat{\rho})$ &  $i \frac{d }{d t} |\rho \rangle\!\rangle = \mathpzc{H}  |\rho \rangle\!\rangle$
\\
\hline
\end{tabular}
\caption{Open quantum system of the Kitaev spin liquid in two equivalent descriptions.
(a) Original density matrix description. The Kitaev honeycomb model is embedded in the Lindblad master equation with the jump operator, $L_\mu = \sqrt{\gamma}\sigma_j^\alpha \sigma_k^\alpha$. 
(b) Doubled state vector description. Vectorization of the Lindblad master equation leads to a closed system defined on the bilayer honeycomb lattice with the non-Hermitian interlayer interaction, $i \gamma ( \sigma_j^\alpha \sigma_k^\alpha \tau_j^\alpha \tau_k^\alpha - 1 )$.
The two descriptions are compared in the table.
}
\label{fig:1}
\end{figure*}

The Lindblad dynamics of the Kitaev spin liquid presents rich physics with a deep connection to anyon condensation, which is an important mechanism for understanding topological transitions between distinct quantum spin liquids~\cite{Bombin2008,Bais2009,Burnell2011,Burnell2018,Schmidt2020,Hwang2024}.
By mapping the Lindblad system to a doubled non-Hermitian system~\cite{Katsura2019,Katsura2019-2}, we uncover that the time evolution from the pure KSL to the mixed-state KSL is actually a dynamical transition from a KSL bilayer product state to a toric code type $\mathbb{Z}_2$ spin liquid state by anyon condensation.

Our study illuminates a new aspect of open quantum systems, i.e., a useful platform for studying topological transitions in quantum spin liquids.
In addition to the Kitaev spin liquid, we study open quantum systems of the toric code model~\cite{Kitaev2003}.
Based on the concrete examples, we elucidate the mechanism of dynamical anyon condensation transitions in quantum spin liquids induced by the Lindblad dynamics.

\section{Kitaev Lindbladian}

We consider an open quantum system of the Kitaev spin liquid coupled to a Markovian environment.
The dynamics of the system is investigated by using the Lindblad master equation~\cite{Lindblad1976,Gorini1976,BreuerPetruccione,Albert2014,Albert2016,Lidar2020,Harrington2022},
\begin{equation}
\frac{d \hat{\rho}}{d t} = \mathcal{L}(\hat{\rho}) 
= 
-i[H,\hat{\rho}] + \sum_{\mu} (L_\mu \hat{\rho} L_\mu^\dagger-\frac{1}{2}\{ L_\mu^\dagger L_\mu, \hat{\rho} \}) ,
\label{eq:LME}
\end{equation}
where the system's density matrix $\hat{\rho}(t)$ evolves in time by the Lindbladian $\mathcal{L}$ composed of the Hamiltonian $H$ and the Lindblad operators $\{L_\mu\}$ describing the environment effects.
The Lindblad equation has the completely positive and trace-preserving property, i.e., ${\rm Tr}\hat{\rho}(t)={\rm const}$. 
For the Hamiltonian, we consider the Kitaev honeycomb model~\cite{Kitaev2006},
\begin{equation}
H
=
K\sum_{\langle jk \rangle_x} \sigma_j^x \sigma_k^x
+
K\sum_{\langle jk \rangle_y} \sigma_j^y \sigma_k^y
+
K\sum_{\langle jk \rangle_z} \sigma_j^z \sigma_k^z,
\end{equation}
where $\sigma^{x,y,z}$ mean the Pauli matrices and $\langle jk \rangle_{x,y,z}$ denote $x,y,z$-bonds of the honeycomb lattice [Fig.~\ref{fig:1}(a)].
For the Lindblad operators generating non-unitary dynamics, we assume the Kitaev bond interactions,
\begin{equation}
L_\mu = \sqrt{\gamma} \sigma_j^\alpha \sigma_k^\alpha,
\end{equation}
with the dissipation strength $\gamma~(\ge0)$.
The Lindblad operator generates decoherence and dissipation effects on the Kitaev spin liquid state by creating a pair of fermion excitations at both sites of the bond $\langle jk \rangle_\alpha$.
The Hamiltonian $H$ and the Lindblad operators $\{L_\mu\}$ both commute with the $\mathbb{Z}_2$ flux operator,
\begin{equation}
\hat{W}_p = 
\sigma_1^z 
\sigma_2^y
\sigma_3^x
\sigma_4^z
\sigma_5^y
\sigma_6^x ,
\end{equation}
i.e., $[H,\hat{W}_p]=[L_\mu,\hat{W}_p]=0$.
Hence, the Kitaev Lindbladian $\mathcal{L}$ conserves the $\mathbb{Z}_2$ flux quantum number.

\begin{figure*}[tb]
\includegraphics[width=\linewidth]{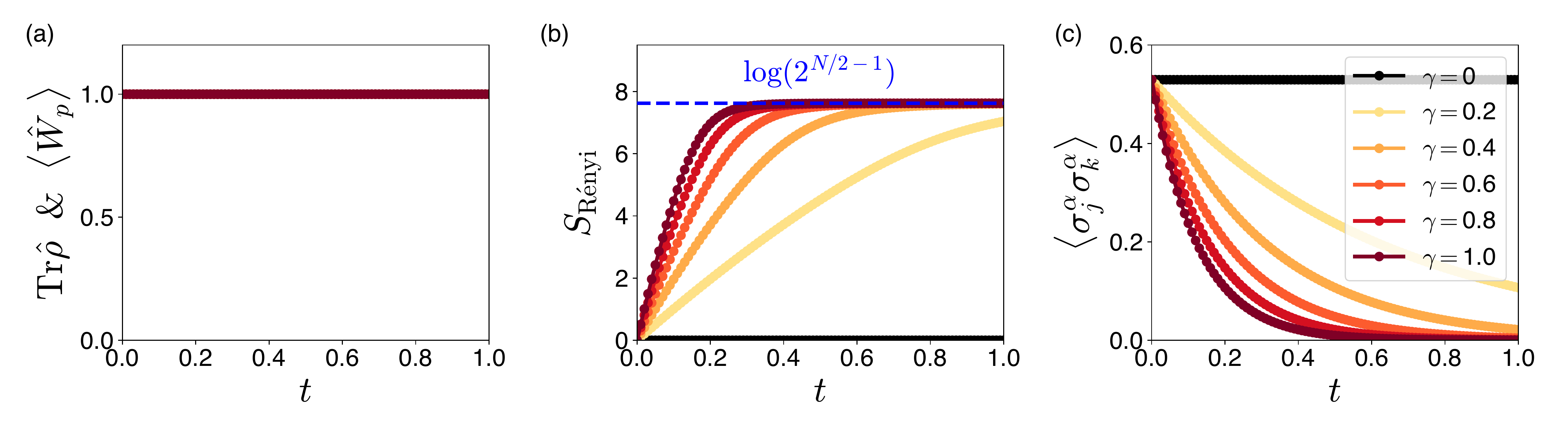}
\caption{Lindblad dynamics of the Kitaev spin liquid.
(a) The trace of the density matrix ${\rm Tr}\hat{\rho(t)}$ and the expectation value of the flux operator $\langle \hat{W}_p \rangle$.
(b) The R\'enyi entropy $S_{\textup{R\'enyi}}$. The dashed line marks the steady state value, $\log(2^{N/2-1})$, where $N=24$ is the number of sites.
(c) The spin-spin correlator $\langle {\sigma}_j^\alpha {\sigma}_k^\alpha \rangle$ for the nearest-neighbor bond $\langle jk\rangle_{\alpha}$.
Different colors denote the results of different dissipation strengths ($\gamma=0,~0.2,~0.4,~0.6,~0.8,~1$). The unitary evolution of the closed system ($\gamma=0$) is shown together for comparison (black).
In all the calculations, the coupling constant is fixed by $K=-1$.}
\label{fig:2}
\end{figure*}

\subsection{Lindblad dynamics of the Kitaev spin liquid}

To understand the dynamics generated by many-body quantum entanglement and the environment effects in an unbiased way, we numerically solve the Lindblad equation by putting the system on a 24-site cluster with periodic boundary condition [Fig.~\ref{fig:1}(a)].
First, we prepare the system in the pure state of the Kitaev spin liquid,
\begin{equation}
 \hat{\rho}(t=0)= | \Psi_{\rm KSL} \rangle \langle \Psi_{\rm KSL} |,
 \label{eq:initial-rho}
\end{equation}
where $| \Psi_{\rm KSL} \rangle$ is the ground state of $H$ which we obtain by exact diagonalization on the 24-site cluster (torus geometry).
The model $H$ has threefold ground state degeneracy, among which we choose the ground state $| \Psi_{\rm KSL} \rangle$ in the sector of the Wilson loop flux $\mathcal{W}_{x,y}=-1$ in our calculations (see Appendix~\ref{appendix:Wilson}).
The time evolution of the density matrix, $\hat{\rho}(t)=e^{t\mathcal{L}}\hat{\rho}(0)$, is computed by using the Krylov subspace methods~\cite{Saad1992,sidje1998expokit}. In the construction of the time evolution operator $e^{t\mathcal{L}}$, we utilize the $\mathbb{Z}_2$ flux quantum numbers of the Kitaev Lindbladian to reduce the size of the Hilbert space.

\begin{figure}[tb]
\includegraphics[width=\linewidth]{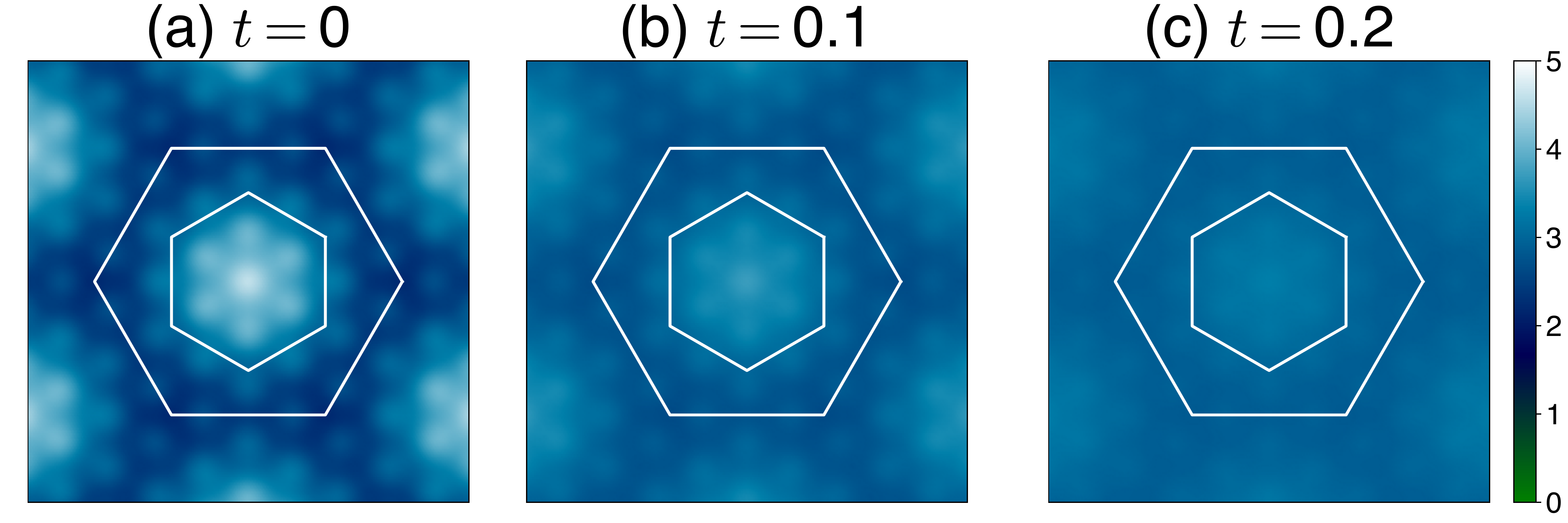}
\caption{Time evolution of the spin structure factor $S({\bf q})$.
The three plots at $t=0,0.1,0.2$ are obtained for the case of $K=-1$ and $\gamma=1$. In each plot, the two hexagons represent the first and second Brillouin zones in momentum space.}
\label{fig:3}
\end{figure}

Figure~\ref{fig:2} shows the calculation results of the Lindblad dynamics.
First, we check the properties ${\rm Tr}\hat{\rho}(t)=1$ and $\langle \hat{W}_p \rangle = {\rm Tr} [ \hat{\rho}(t) \hat{W}_p ]=1$ in Fig.~\ref{fig:2}(a).
Next, we consider the R\'enyi entropy,
\begin{equation}
S_{\textup{R\'enyi}}=-{\rm log}[{\rm Tr} \hat{\rho}^2(t)],
\end{equation}
which tells us how the system's purity changes by the environment effects.
As shown in Fig.~\ref{fig:2}(b), the R\'enyi entropy gradually increases converging to the constant value: $S_{\textup{R\'enyi}}=\log(2^{N/2-1})$ when $t\rightarrow\infty$ ($N=24$ is the number of sites). 
The larger the dissipation strength $\gamma$ is, the faster the convergence is reached.
The constant value reveals the size of the subspace ($2^{N/2-1}=2,048$) that the system belongs to, i.e., the zero flux sector with the Wilson loop flux $\mathcal{W}_{x,y}=-1$ (see Appendix~\ref{appendix:Wilson}).

An interesting behavior appears in the spin-spin correlation $\langle {\sigma}_j^\alpha {\sigma}_k^\alpha \rangle = {\rm Tr} [ \hat{\rho}(t) {\sigma}_j^\alpha {\sigma}_k^\alpha ]$. It monotonically decreases and vanishes in the long time limit: $\langle {\sigma}_j^\alpha {\sigma}_k^\alpha \rangle=0$ when $t\rightarrow\infty$ [Fig.~\ref{fig:2}(c)].
The system is still in a quantum spin liquid phase since it stays in the zero flux sector. Nonetheless, the nearest neighbor bond spin-spin correlation disappears in the long time limit.
This peculiar property is also reflected in the spin structure factor,
\begin{equation}
S({\bf q}) = \frac{1}{N} \sum_{j,k} e^{i{\bf q}\cdot({\bf r}_j-{\bf r}_k)}  \langle {\boldsymbol{\sigma}}_j \cdot {\boldsymbol{\sigma}}_k \rangle ,
\end{equation}
which becomes flat in momentum space in the long time limit: $S({\bf q})=3$ when $t\rightarrow\infty$ (see Fig.~\ref{fig:3}).

\subsection{Exact solution of the steady state}

\begin{figure}[tb]
\includegraphics[width=\linewidth]{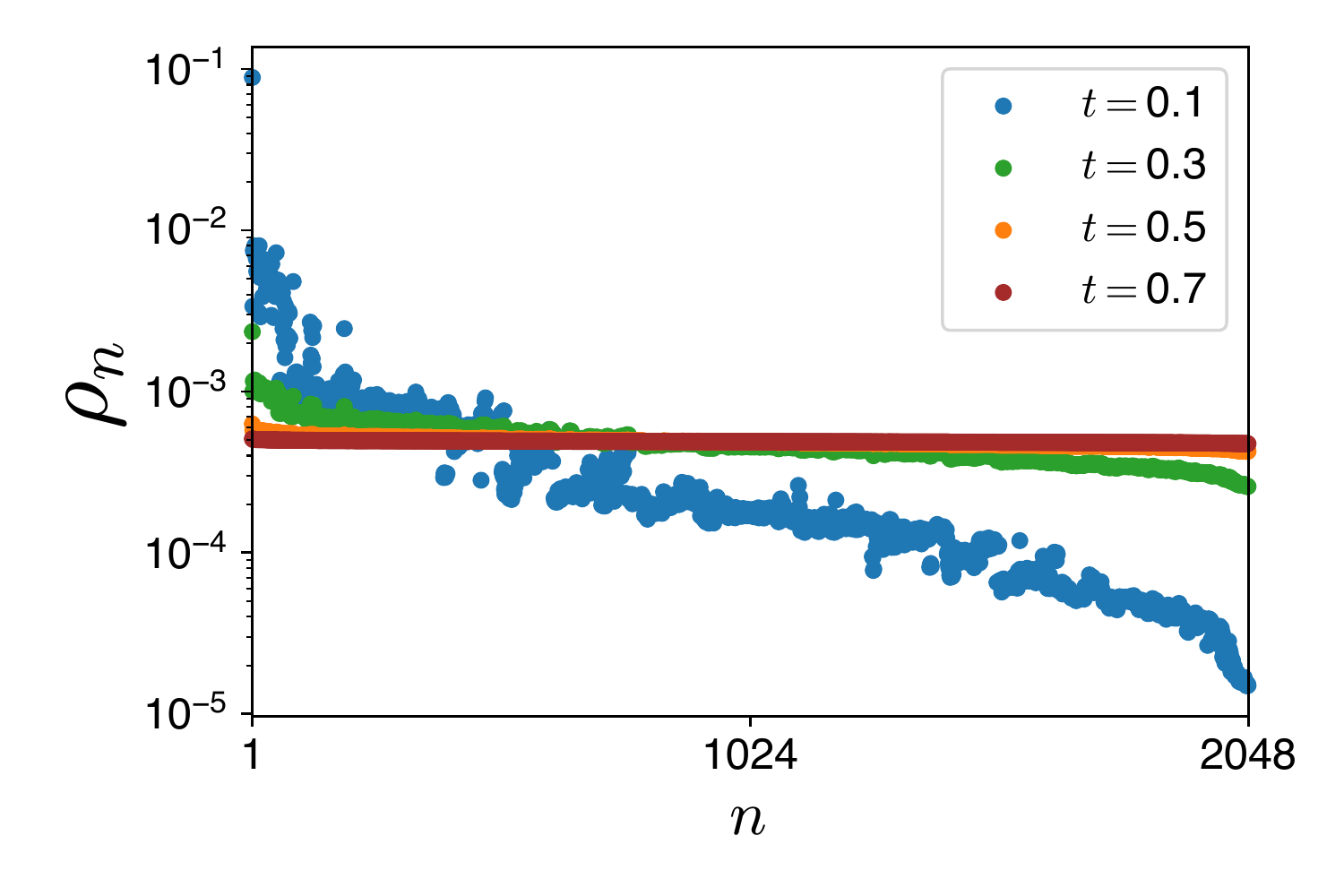}
\caption{Decomposition of the density matrix.
Numerically obtained density matrices at different times ($t=0.1,0.3,0.5,0.7$) are decomposed by $\rho_n=\langle \Psi_n | \hat{\rho}(t) | \Psi_n \rangle$. At $t=0.7$, the components become almost constant, $\rho_n=2^{-N/2+1}\simeq4.9\times10^{-4}$ ($N=24$). The results are obtained with the coupling constants, $K=-1$ and $\gamma=1$.}
\label{fig:4}
\end{figure}

The numerical approach establishes the existence of the steady state characterized by the properties, $\langle \hat{W}_p \rangle = 1$, $S_{\textup{R\'enyi}}=\log(2^{N/2-1})$, and $\langle {\sigma}_j^\alpha {\sigma}_k^\alpha \rangle=0$.
We find the exact solution for the steady state:
\begin{eqnarray}
\hat{\rho}_{\rm MSKSL}
&=&
\frac{1}{2^{N+1}}\prod_{l=x,y} (1-\hat{\mathcal{W}}_l)\prod_p (1+\hat{W}_p)
\nonumber\\
&=&
\sum_n \rho_n | \Psi_n \rangle \langle \Psi_n |.
\label{eq:rho-NKSL}
\end{eqnarray}
The steady state density matrix is simply given by the projection operator into the zero flux sector with the Wilson loop flux $\mathcal{W}_{x,y}=-1$.
Hence, it is diagonal in the eigenbasis $\{ | \Psi_n \rangle \}$ of $H$ and $\hat{W}_p$ with the weight
\begin{equation}
\rho_n
=
\left\{
\begin{array}{cc}
\frac{1}{2^{N/2-1}} & (W_p=+1~\&~\mathcal{W}_{x,y}=-1)
\\
0 & ({\rm otherwise})
\end{array}
\right. .
\label{eq:rho_n}
\end{equation}
This equal weight property is indeed identified in our numerical calculations shown in Fig.~\ref{fig:4}.
One can check that $\mathcal{L}(\hat{\rho}_{\rm MSKSL})=0$ along with the three properties mentioned earlier. This is the maximally mixed state in the zero-flux sector with equal weight,
which we call ``mixed-state Kitaev spin liquid (MSKSL)''.

\section{Mapping to the non-Hermitian bilayer model}

Lindblad system allows an exact mapping to a doubled system described by a non-Hermitian Hamiltonian~\cite{Katsura2019,Katsura2019-2}.
The mapping is conducted by the vectorization of the density matrix called the Choi-Jamio{\l}kowski  isomorphism~\cite{Jamiolkowski1972,Choi1975}:
\begin{eqnarray}
\hat{\rho}
=
\sum_{m,n} \rho_{mn} |m\rangle \langle n|
~~\Rightarrow~~
|\rho\rangle\!\rangle
=
\sum_{m,n} \rho_{mn} |m\rangle \otimes |n\rangle,~~~
\end{eqnarray}
where the bra of the density matrix $\hat{\rho}$ is turned into a ket of the resulting state vector $|\rho\rangle\!\rangle$.
The system is now effectively doubled with an additional copy of the original Hilbert space.
By the vectorization,
the Lindblad master equation takes the form of the Schr\"odinger equation
\begin{eqnarray}
i \frac{d |\rho\rangle\!\rangle}{d t} = \mathpzc{H}  |\rho\rangle\!\rangle 
\label{eq:non-Hermitian-Schrodinger}
\end{eqnarray}
with the non-Hermitian Hamiltonian
\begin{eqnarray}
\mathpzc{H} 
&=& 
H \otimes I - I \otimes H^T
\\
&+&i\sum_{\mu} 
\left[ 
L_\mu \otimes L_\mu^* 
-
\frac{1}{2}
(L_\mu^\dagger L_\mu) \otimes I 
-
\frac{1}{2}
I \otimes (L_\mu^\dagger L_\mu)^* 
\right].
\nonumber
\end{eqnarray}

In our case, the non-Hermitian Hamiltonian is given by
\begin{eqnarray}
\mathpzc{H}
&=&
K\sum_{\langle jk \rangle_\alpha}^{\textup{upper layer}} \sigma_j^\alpha \sigma_k^\alpha
-
K\sum_{\langle jk \rangle_\alpha}^{\textup{lower layer}} \tau_j^\alpha \tau_k^\alpha
\nonumber\\
&+&
i \gamma \sum_{\langle jk \rangle_\alpha}^{\textup{inter-layer}} ( \sigma_j^\alpha \sigma_k^\alpha \tau_j^\alpha \tau_k^\alpha - 1 ),
\end{eqnarray}
where $\sigma$ \& $\tau$ are Pauli spin operators acting on the original kets and copied kets; $\sigma_j^\alpha \tau_k^\beta (| m \rangle \otimes | n \rangle) = (\sigma_j^\alpha | m \rangle) \otimes (\tau_k^\beta  | n \rangle)$.
We now have a bilayer spin model with the intralayer Kitaev interactions ($K$ \& $-K$) and the interlayer interactions ($i\gamma$) [Fig.~\ref{fig:1}(b)]. Note that the interlayer interactions are non-Hermitian originating from the environment effects. This leads to non-unitary time evolution of the state vector, $ |\rho(t)\rangle\!\rangle = e^{-i t \mathpzc{H}}  |\rho(0)\rangle\!\rangle $, which is identified from the relationship $\langle\!\langle \rho | \rho \rangle\!\rangle=\exp(-S_{\textup{R\'enyi}})$ and Fig.~\ref{fig:2}(b).

The non-Hermitian bilayer model has two types of $\mathbb{Z}_2$ flux operators, 
$
\hat{W}_p 
$
and
$
\hat{Z}_p = 
\tau_1^z 
\tau_2^y
\tau_3^x
\tau_4^z
\tau_5^y
\tau_6^x ,
$
commuting with the Hamiltonian ($[\mathpzc{H},\hat{W}_p]=[\mathpzc{H},\hat{Z}_p]=[\hat{W}_p,\hat{Z}_{p'}]=0$).
This model can be viewed as a non-Hermitian analog of the bilayer spin model in Ref.~\cite{Hwang2024}.

\subsection{Steady state degeneracy}

The bilayer model approach enables us to discover exact steady state solutions and their extensive degeneracy. We find the steady state condition,
\begin{equation}
\sigma_j^\alpha \sigma_k^\alpha \tau_j^\alpha \tau_k^\alpha |\rho\rangle\!\rangle = |\rho\rangle\!\rangle
~~~\Longleftrightarrow~~~
\mathpzc{H} |\rho\rangle\!\rangle = 0.
\label{eq:steady-state-condition}
\end{equation}
Any state that satisfies the above local constraint is basically a steady state.
For instance, we may consider the singlet product state,
\begin{equation}
|\lambda\rangle\!\rangle = \otimes_j | s \rangle_j,
\end{equation}
where $| s \rangle=\frac{1}{\sqrt{2}}(|\uparrow\rangle \otimes |\downarrow \rangle - | \downarrow\rangle \otimes |\uparrow \rangle)$ is the spin-singlet of $\sigma$ and $\tau$ spins.
One can check that $\sigma_j^\alpha \sigma_k^\alpha \tau_j^\alpha \tau_k^\alpha |\lambda\rangle\!\rangle = |\lambda\rangle\!\rangle$.
From the singlet product state, we may generate further steady states as follows. 
\begin{equation}
| \rho_{\rm ss} \{n\} \rangle\!\rangle= \prod_{l=x,y}(\hat{\mathcal{W}}_l)^{n_l} \prod_p (\hat{W}_p)^{n_p} | \lambda \rangle\!\rangle,
\end{equation}
where $n_l=0,1~(l={x,y})$, and $n_p=0,1$ at each plaquette $p$.
Instead of the states, we consider the $\mathbb{Z}_2$ flux eigenstates,
\begin{equation}
| \rho_{\rm ss} \{ w \}\rangle\!\rangle
=
 \mathcal{N} \prod_{l=x,y}(1+w_l\hat{\mathcal{W}}_l)
 \prod_p (1+w_p \hat{W}_p) | \lambda \rangle\!\rangle ,
\end{equation}
where $\mathcal{N}$ is a normalization constant, and $w_p(=\pm1)$ \& $w_{l}(=\pm1)$ specify the flux sector including the Wilson loop flux.
By counting all these states, we find the degeneracy of the steady state manifold, or steady state degeneracy (SSD):
\begin{equation}
{\rm SSD}=2^2\times2^{N/2-1}.
\end{equation}
The first factor accounts for the four Wilson loop flux sectors ($w_{x}=\pm1~\&~w_{y}=\pm1$) while the second one counts the number of possible flux sectors ($w_p=\pm1$) on torus geometry.

By applying the inverse vectorization ($|s\rangle \Rightarrow i\sigma^y/\sqrt{2}$) to $| \rho_{\rm ss} \{ w \}\rangle\!\rangle$, we obtain the density matrices of the steady states:
\begin{eqnarray}
\hat{\rho}_{\rm ss}\{ w \} 
=
\mathcal{N} \prod_{l=x,y}(1+w_l\hat{\mathcal{W}}_l)
\prod_p (1+w_p \hat{W}_p) \prod_j \frac{i\sigma_j^y}{\sqrt{2}} .~~
\label{eq:rho_ss}
\end{eqnarray}
On torus geometry, $\prod_j \frac{i\sigma_j^y}{\sqrt{2}}$ is equivalent to a product of flux operators thus can be absorbed into the term $\prod_p (1+w_p \hat{W}_p)$; see Appendix~\ref{appendix:exact_sol}.
If we focus on the flux sector ($w_p=+1,w_l=-1$), we obtain the steady state in Eq.~(\ref{eq:rho-NKSL}).

\begin{figure}[tb]
\includegraphics[width=\linewidth]{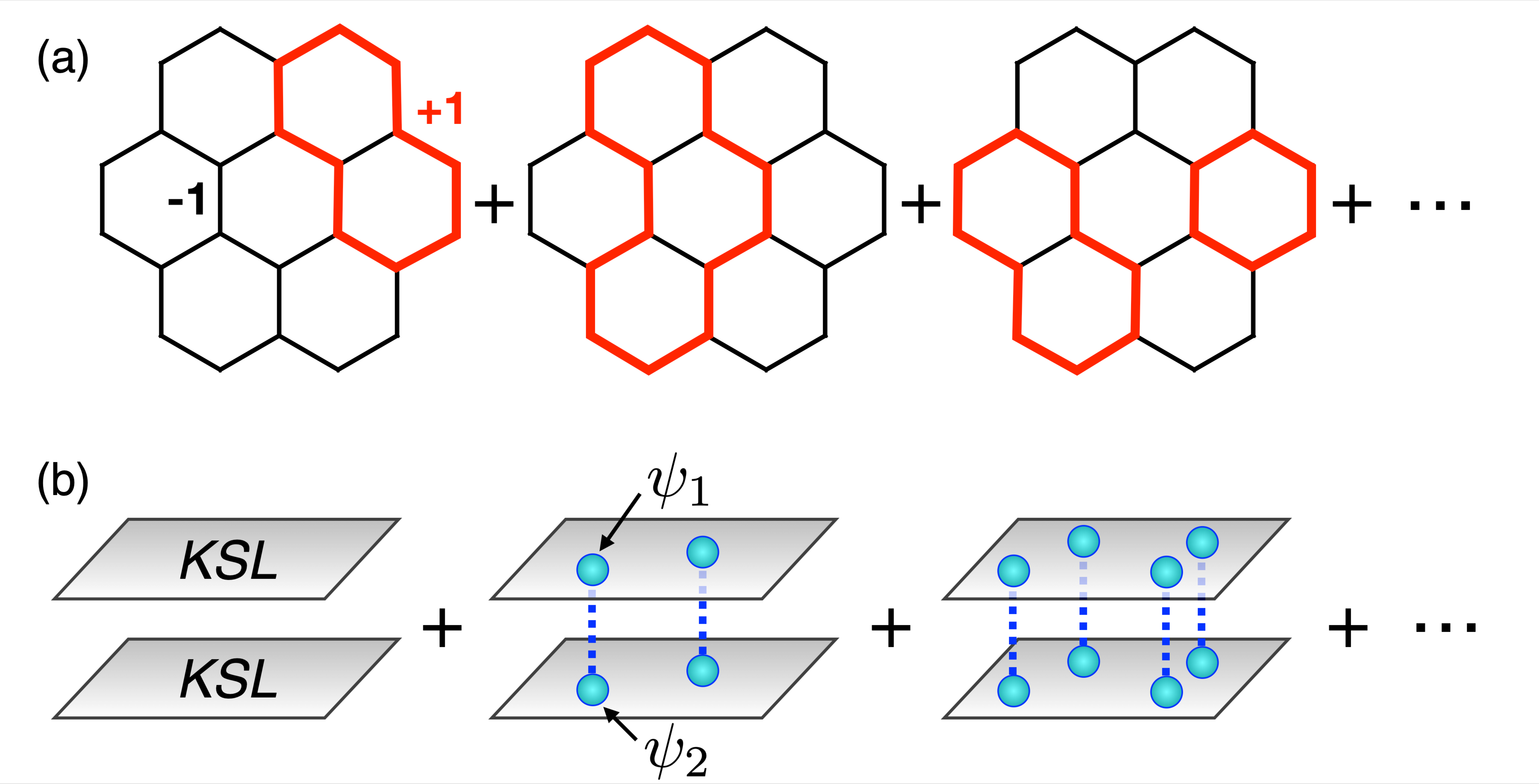}
\caption{Two equivalent representations of the $\mathbb{Z}_2$ spin liquid steady state $|\rho(t\rightarrow\infty)\rangle\!\rangle$.
(a) Toric code state representation. Black lines denote the $\mathbb{Z}_2$ variables $\{X_{jk}=-1\}$. The sign-flipped $\mathbb{Z}_2$ variables $\{X_{jk}=+1\}$ are depicted by red lines.
(b) Anyon condensation representation. The condensed anyons are fermion pairs across the layers (blue balls connected by a dashed line).}
\label{fig:5}
\end{figure}

\subsection{Dynamical anyon condensation}

A fascinating physics is hidden in the Lindblad dynamics of Fig.~\ref{fig:2}, which we unveil by using the bilayer description.
The initial state is the KSL bilayer product state, $|\rho(t=0)\rangle\!\rangle = |\Psi_{\rm KSL}\rangle\otimes|\Psi_{\rm KSL}\rangle$,
and the steady state is given by $|\rho(t\rightarrow\infty)\rangle\!\rangle = |\rho_{\rm MSKSL}\rangle\!\rangle $, i.e., 
\begin{equation}
|\rho(t\rightarrow\infty)\rangle\!\rangle 
=
\frac{1}{2^{N/2+1}} 
\prod_{l=x,y}(1-\hat{\mathcal{W}}_l)  
\prod_p (1+\hat{W}_p) | \lambda \rangle\!\rangle .
\nonumber
\end{equation}
We find that this steady state is nothing but a $\mathbb{Z}_2$ spin liquid.
Therefore, the time evolution from $|\rho(t=0)\rangle\!\rangle$ to $|\rho(t\rightarrow\infty)\rangle\!\rangle$ is a dynamical transition from the KSL$\times$KSL state to a $\mathbb{Z}_2$ spin liquid state.

How does the state $|\rho(t\rightarrow\infty)\rangle\!\rangle$ represent a $\mathbb{Z}_2$ spin liquid?
First, we note that $| \lambda \rangle\!\rangle$ satisfies the condition $\phi_j^x =\phi_j^y =\phi_j^z=-1$ at every site, where $\phi_j^\alpha\equiv\sigma_j^\alpha\tau_j^\alpha$~\cite{Hwang2024}.
If a flux operator $\hat{W}_p$ acts on $| \lambda \rangle\!\rangle$, it flips the sign of $\phi_j^\alpha$ ($-1\rightarrow+1$) along the bonds of the plaquette $p$.
Using these properties, we assign a $\mathbb{Z}_2$ variable $X_{jk}(\equiv\phi_j^\alpha=\phi_k^\alpha=\pm1)$ to each bond $\langle jk \rangle_\alpha$.
Then, we immediately recognize that the state $\prod_p (1+\hat{W}_p) | \lambda \rangle\!\rangle$ is equivalent with the $\mathbb{Z}_2$ spin liquid state of the toric code model~\cite{Kitaev2003}; see Fig.~\ref{fig:5}(a).

As already shown in Eqs.~(\ref{eq:rho-NKSL}) and (\ref{eq:rho_n}), the steady state can be represented by
\begin{equation}
|\rho(t\rightarrow\infty)\rangle\!\rangle 
= 
\sum_n \rho_n | \Psi_n \rangle \otimes | \Psi_n \rangle ,
\nonumber
\end{equation}
where $\{| \Psi_n \rangle\}$ are the states of the zero flux sector including the KSL ground state and excited states. Note that each state has an even number of fermion excitations (without any flux excitations).
The state $|\rho(t\rightarrow\infty)\rangle\!\rangle $ is an equal weight superposition of all the bilayer product states $\{| \Psi_n \rangle \otimes | \Psi_n \rangle\}$.
This implies that $|\rho(t\rightarrow\infty)\rangle\!\rangle$ is the $\mathbb{Z}_2$ spin liquid state that emerges from the KSL$\times$KSL state by condensing fermion pairs between the two layers [Fig.~\ref{fig:5}(b)].

\begin{figure}[tb]
\includegraphics[width=\linewidth]{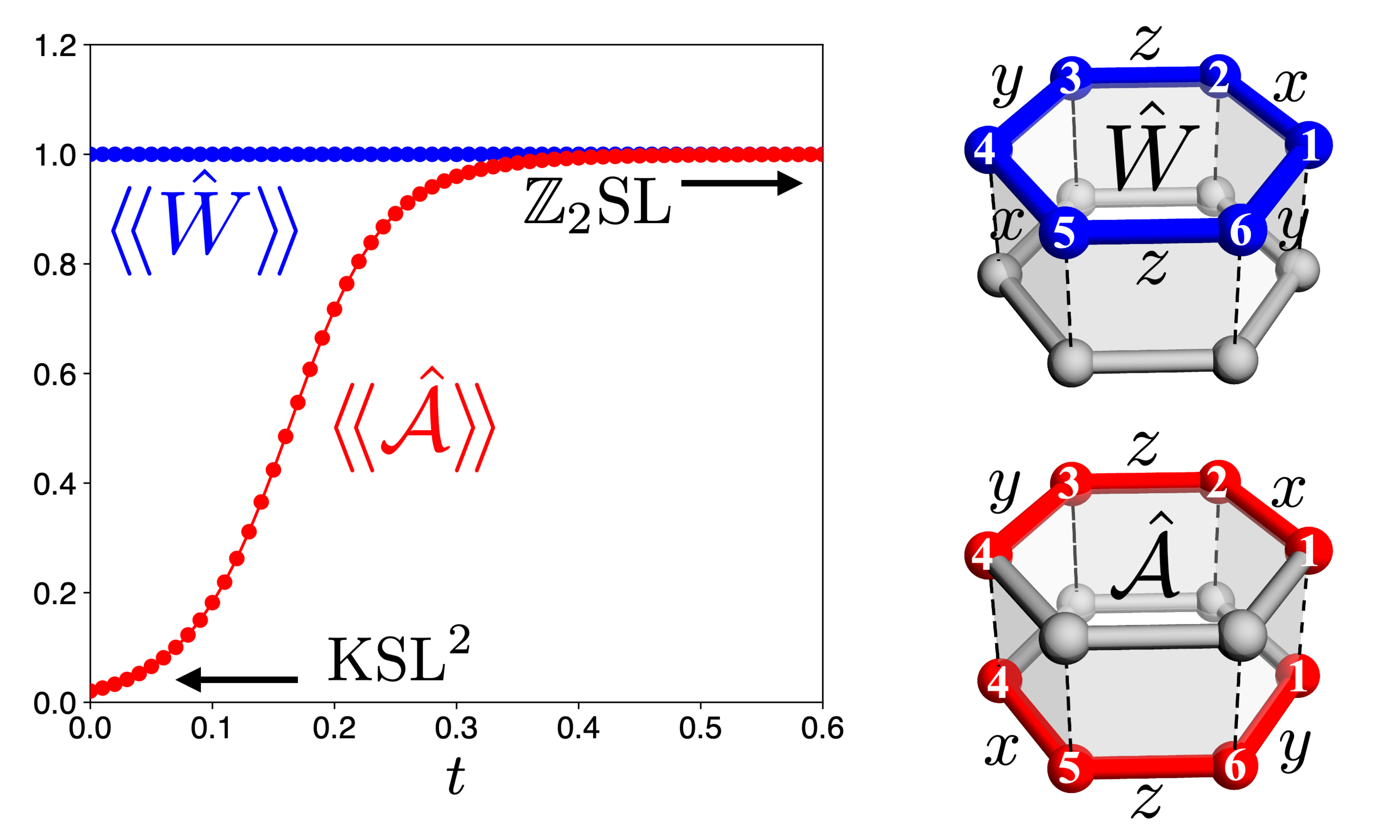}
\caption{Dynamical anyon condensation.
Red points denote the expectation value $\langle\!\langle \hat{\mathcal{A}} \rangle\!\rangle $ as a function of time $t$.
The flux operator expectation value $\langle\!\langle \hat{W} \rangle\!\rangle$ is plotted together for comparison (blue).
In the calculations, the coupling constants are fixed by $K=-1$ and $\gamma=1$.}
\label{fig:6}
\end{figure}

To confirm the fermion pair condensation, we consider the loop operator
\begin{equation}
\hat{\mathcal{A}}=(\tau_1^y\tau_6^y)(\tau_6^z\tau_5^z)(\tau_5^x\tau_4^x)(\sigma_4^y\sigma_3^y)(\sigma_3^z\sigma_2^z)(\sigma_2^x\sigma_1^x) ,
\label{eq:loop-op}
\end{equation}
which measures whether fermion excitations of the KSL can move between the two layers, i.e., the fermion pair condensation between the layers~\cite{Hwang2024}.
We indeed find that the expectation value $\langle\!\langle \hat{\mathcal{A}} \rangle\!\rangle \equiv \langle\!\langle {\rho} | \hat{\mathcal{A}} | {\rho} \rangle\!\rangle/\langle\!\langle {\rho} | {\rho} \rangle\!\rangle$ increases from almost zero ($t=0$) to one ($t\rightarrow\infty$); see Fig.~\ref{fig:6}.
Furthermore, we explicitly check the steady state condition, $\langle\!\langle \sigma_j^\alpha \sigma_k^\alpha \tau_j^\alpha \tau_k^\alpha \rangle\!\rangle=1$, and entanglement entropy between the two layers in the steady state limit (Appendix~\ref{appendix:bilayer}).
All these results confirm the dynamical anyon condensation from the KSL$\times$KSL state to the $\mathbb{Z}_2$ spin liquid state.

\section{Discussion}

This work establishes that the $\mathbb{Z}_2$ toric code (TC) spin liquid is equivalent with an anyon condensed phase of the Kitaev spin liquid bilayer. Here, the condensed anyons are the fermion pairs $\psi_1\boxtimes\psi_2$ ($\psi_{1},\psi_{2}$: fermion excitations in the upper and lower layer KSLs, respectively).
\begin{equation}
\mathcal{L}_{\rm KSL}: {\rm KSL}\times{\rm KSL} \xrightarrow[]{\langle \psi_1\boxtimes\psi_2 \rangle \ne 0} \mathbb{Z}_2{\rm TC}
\end{equation}
We showed this by using the Lindblad dynamics of the KSL and analyzing the exact solution of the mixed-state KSL via the Choi-Jamio{\l}kowski isomorphism.
The results summarized in Eq.~(\ref{eq:rho-NKSL}) and Fig.~\ref{fig:5} elucidate the anyon condensation transition in a simple, exact, but nontrivial way.


The open quantum system of the Kitaev spin liquid reveals a general mechanism for inducing anyon condensations in quantum spin liquids by the Lindblad dynamics.
A quantum spin liquid is usually realized by preserving a certain type of local constraints (e.g., flux quantum number in the Kitaev spin liquid). The quantum spin liquid is embedded in a Lindbladian that preserves the local constraints. Then, the steady state should appear in the form of a maximally mixed state within the subspace defined by the constraints. In the bilayer description, the maximally mixed state corresponds to a new quantum spin liquid emerging from the initial-state spin liquid bilayer by anyon condensation. The key point is that the {\it maximal mixing} in the original system corresponds to the {\it anyon condensation} in the doubled system.


The mechanism applies to general quantum spin liquids.
We take one more example: an open quantum system of the toric code model~\cite{Kitaev2003}.
The toric code system $H_{\rm TC}=-J_{A}\sum_{s} A_s - J_{B}\sum_{p} B_p$~(where $J_{A,B}>0$, $A_s=\prod_{j\in s}\sigma_j^x$ and $B_p=\prod_{j\in p}\sigma_j^z$) is embedded in the Lindbladian with the jump operator $L_{\mu}=\sqrt{\gamma} \sigma_j^x$.
Note that the jump operator creates a pair of $m$-particles ($B_p=-1$) without touching the $e$-particle sector.
In this case, the Lindblad equation [Eq.~(\ref{eq:LME})] is exactly solved in the steady state limit.
The $\mathbb{Z}_2$ toric code spin liquid state $| A_s=1~\&~B_p=1\rangle$ evolves in time to the exact steady state 
\begin{equation}
\hat{\rho}_{\textup{ss}} \propto \prod_s \frac{1+A_s}{2},
\label{eq:TC-steady-state}
\end{equation}
which is the maximally mixed state of all possible $m$-particle excitations (with no $e$-particles). 
In the doubled system description, the maximally mixedness corresponds to the condensation of $m$-particle pairs, $m_1\boxtimes m_2$, yielding another $\mathbb{Z}_2$ toric code spin liquid in the steady state $|\rho_{\textup{ss}}\rangle\!\rangle$.
\begin{equation}
\mathcal{L}_{\rm TC}: \mathbb{Z}_2{\rm TC}\times\mathbb{Z}_2{\rm TC} \xrightarrow[]{\langle m_1\boxtimes m_2 \rangle \ne 0} \mathbb{Z}_2{\rm TC}
\label{eq:TC-DAC}
\end{equation}
One can condense $e$-particles ($A_s=-1$) instead by considering the jump operator $L_{\mu}=\sqrt{\gamma} \sigma_j^z$.
Qualitatively same results are obtained in this case due to the duality between the $e$- and $m$-particles. See Appendix~\ref{appendix:toric_code} for more details.


We make some comments on related works to our study.
Lindblad open quantum systems of the KSL have been studied in a different setup from ours in Ref.~\cite{Yang2021}; they assumed the jump operator $L_{\mu}=\sigma_j^\alpha+\sigma_k^\alpha$ and no occurrence of quantum jump [accordingly, they ignored the term $L_\mu \hat{\rho} L_\mu^\dagger$ in the master equation, Eq.~(\ref{eq:LME})].
This work shows that the non-Hermitian physics without quantum jump leads to interesting spin liquids featured with exceptional points, fermi arcs, and non-Hermitian skin effects.
Such exceptional spin liquids are only maintained in the absence of any quantum jump event. By contrast, our mixed-state spin liquids appear in the steady state limit as an outcome of quantum jump events.

Recently, decoherence-induced transitions and mixed-state topological orders are widely investigated in various aspects.
The same anyon condensation physics in Eq.~(\ref{eq:TC-DAC}) was shown to arise in the toric code subject to local decoherence channels~\cite{Lee2023,Bao2023}. 
For the intrinsic characterizations of mixed-state topological orders, several information-theoretical diagnostics have been proposed and examined such as quantum relative entropy, coherent information, and topological entanglement negativity~\cite{Fan2024,Wang2023}.
General frameworks to characterize/classify mixed-state topological orders are also being developed very recently~\cite{Sohal2024,Ellison2024}.
Our mixed-state spin liquids appearing from the Lindblad dynamics provide concrete realizations of mixed-state topological orders.
Exact solutions of our mixed-state spin liquids are expected to help to sharpen our understanding on mixed-state topological orders.

To summarize, we proposed a new platform for anyon condensation topological transitions, i.e.,  open quantum spin liquids.
The main idea is to induce a mixed-state quantum spin liquid in the steady state limit by using the Lindblad dynamics, which gives rise to an anyon-condensed phase in the corresponding doubled system.
This framework can be applicable to more generic systems with topological orders.

\acknowledgements

I thank Jong~Yeon~Lee, Gil~Young~Cho, and Jung~Hoon~Han for valuable discussions.
This work was supported by KIAS Individual Grants (No.~PG071402~\&~PG071403) at Korea Institute for Advanced Study (KIAS). 
Computations were performed on clusters at the Center for Advanced Computation (CAC) of KIAS.

\appendix

\begin{figure*}[tb]
\includegraphics[width=\linewidth]{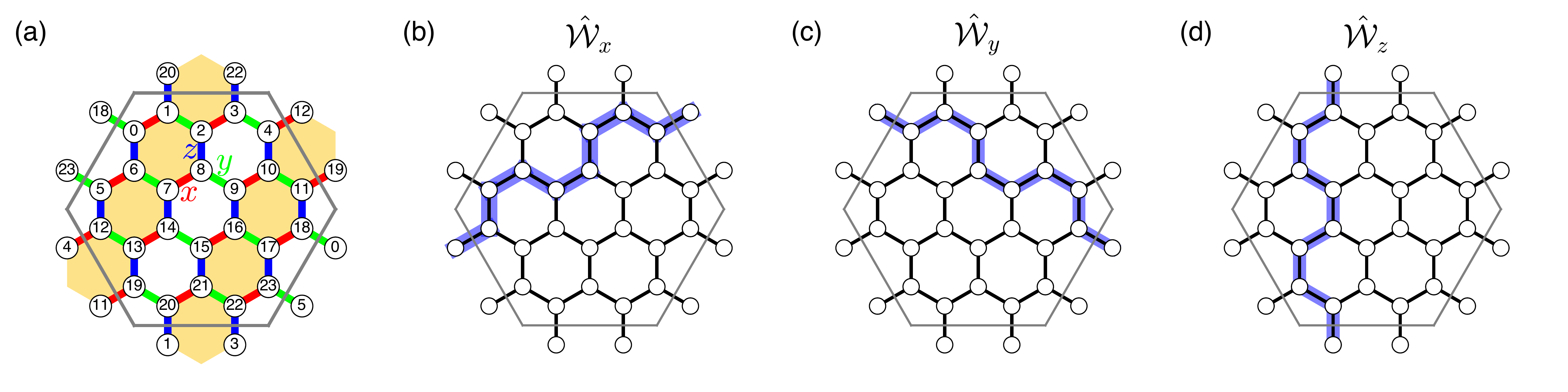}
\caption{The 24-site cluster and the Wilson loop operators}
\label{fig:S1}
\end{figure*}

\section{Wilson loop operators and the reduced Hilbert space\label{appendix:Wilson}}

To reduce the size of the Hilbert space in the actual calculations, we consider two different Wilson loop flux operators $\hat{\mathcal{W}}_{x,y}$ (in addition to the hexagon flux operators $\{\hat{W}_p\}$).
The Wilson loop flux operators are defined by generalizing the hexagon flux operators to non-contractible loops of the system.
The 24-site cluster used in our calculations is shown in Fig.~\ref{fig:S1}(a);
a periodic boundary condition is imposed across the gray hexagon.
Along the non-contractible loop in Fig.~\ref{fig:S1}(b), $\hat{\mathcal{W}}_x$ is defined by
\begin{eqnarray}
\hat{\mathcal{W}}_x 
&=&
 (\sigma_{12}^z \sigma_5^z) (\sigma_5^x \sigma_6^x) (\sigma_6^y \sigma_7^y) (\sigma_7^x \sigma_8^x) 
\nonumber\\
&&
 (\sigma_8^z \sigma_2^z) (\sigma_2^x \sigma_3^x) (\sigma_3^y \sigma_4^y) (\sigma_4^x \sigma_{12}^x)
\\
&=&
- \sigma_{12}^y \sigma_5^y \sigma_6^z \sigma_7^z
 \sigma_8^y \sigma_2^y \sigma_3^z \sigma_4^z .
\end{eqnarray}
This is nothing but the product of the Kitaev terms along the non-contractible loop.
Repeating for other possible non-contractible loops [Fig.~\ref{fig:S1}(c),(d)],
we can define other Wilson loop flux operators:
\begin{eqnarray}
\hat{\mathcal{W}}_y 
=
- \sigma_0^z \sigma_1^z \sigma_2^x \sigma_8^x
 \sigma_9^z \sigma_{10}^z \sigma_{11}^x \sigma_{18}^x , 
\end{eqnarray}
\begin{eqnarray}
\hat{\mathcal{W}}_z 
=
- \sigma_1^y \sigma_0^y \sigma_6^x \sigma_7^x
 \sigma_{14}^y \sigma_{13}^y \sigma_{19}^x \sigma_{20}^x .
\end{eqnarray}
All these Wilson loop flux operators commute with the Lindbladian: $[H,\hat{\mathcal{W}}_{l}]=[L_{\mu},\hat{\mathcal{W}}_{l}]=0$ ($l=x,y,z$).
Note that the three operators are not independent of each other in the zero flux sector ($W_p=+1$) due to the relationship $\hat{\mathcal{W}}_x \hat{\mathcal{W}}_y \hat{\mathcal{W}}_z = +1$.
Among the three operators, we use $\hat{\mathcal{W}}_x$ and $\hat{\mathcal{W}}_y$.
Thanks to the flux quantum numbers of $\hat{\mathcal{W}}_{x,y}$ and $\{\hat{W}_p\}$, the size of the Hilbert space in each flux sector is significantly reduced to
\begin{equation}
\frac{2^N}{2^2\cdot2^{N/2-1}}=2^{N/2-1}.
\end{equation}

By solving the Hamiltonian $H$ by exact diagonalization, we obtain threefold degenerate ground states $\{|\psi_{\rm KSL}^{\rm I} \rangle, |\psi_{\rm KSL}^{\rm II} \rangle, |\psi_{\rm KSL}^{\rm III} \rangle \}$ which are distinguished by different eigenvalues of the Wilson loop flux operators.
\begin{equation}
\begin{array}{c|ccc|c}
\hline
& \mathcal{W}_x & \mathcal{W}_y & \mathcal{W}_z & W_p 
\\
\hline
|\psi_{\rm KSL}^{\rm I} \rangle & +1 & -1 & -1 & +1
\\
|\psi_{\rm KSL}^{\rm II} \rangle & -1 & +1 & -1 & +1
\\
|\psi_{\rm KSL}^{\rm III} \rangle & -1 & -1 & +1 & +1
\\
\hline
\end{array}
\end{equation}
In our calculations, the state $|\psi_{\rm KSL}^{\rm III}\rangle$ is chosen for the initial state of the Lindblad dynamics [Eq. (\ref{eq:initial-rho})].

\section{Exact steady state solution\label{appendix:exact_sol}}

Here we show the relationship,
\begin{equation}
\prod_j \frac{i\sigma_j^y}{\sqrt{2}} \propto \prod_{p\in \mathcal{P}} \hat{W}_p,
\label{eq:sigmay-to-flux}
\end{equation}
where $\mathcal{P}$ means the set of the hexagon plaquettes marked in Fig.~\ref{fig:S1}(a).
The hexagon plaquettes cover a half of the cluster in the stripe pattern perpendicular to the $y$-bond direction.
In this case, one can find that 
\begin{equation}
\prod_{p\in \mathcal{P}} \hat{W}_p \propto \prod_{\langle jk \rangle_x} \sigma_j^x \sigma_k^x \prod_{\langle jk \rangle_z} \sigma_j^z \sigma_k^z \propto \prod_{j} i \sigma_j^y .
\end{equation}
From this, we identify the relationship, Eq.~(\ref{eq:sigmay-to-flux}).
If we apply this to Eq.~(\ref{eq:rho_ss}) together with $\mathcal{W}_{x,y}=-1$, $W_p=+1$ and appropriate normalization, we finally obtain the solution in Eq.~(\ref{eq:rho-NKSL}).

The relationship of Eq.~(\ref{eq:sigmay-to-flux}) is valid on generic clusters if the periodic boundary condition allows covering the hexagon plaquettes in the stripy pattern as shown in Fig.~\ref{fig:S1}(a).

\begin{figure}[b]
\includegraphics[width=\linewidth]{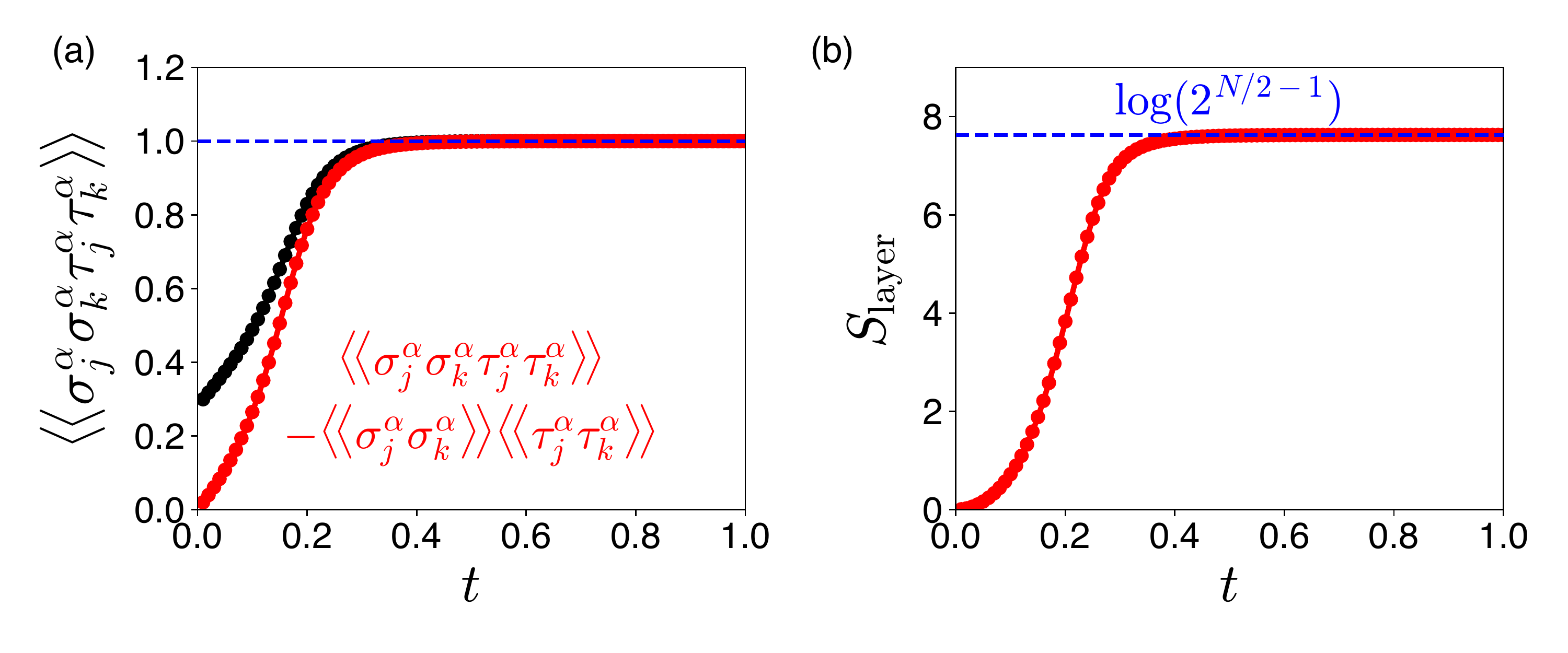}
\caption{Numerical results of the non-Hermitian Kitaev bilayer.
(a) The steady state condition $\sigma_j^\alpha \sigma_k^\alpha \tau_j^\alpha \tau_k^\alpha=1$. 
(b) The entanglement entropy $S_{\rm layer}$.}
\label{fig:S2}
\end{figure}

\section{Non-Hermitian Kitaev bilayer\label{appendix:bilayer}}

To get convinced about the nature of the steady state, 
we have examined other quantities of the non-Hermitian Kitaev bilayer.
We checked the steady state condition [Eq.~(\ref{eq:steady-state-condition})] by computing the expectation value,
\begin{equation}
\langle\!\langle \sigma_j^\alpha \sigma_k^\alpha \tau_j^\alpha \tau_k^\alpha \rangle\!\rangle
=
\frac{\langle\!\langle {\rho} | \sigma_j^\alpha \sigma_k^\alpha \tau_j^\alpha \tau_k^\alpha | {\rho} \rangle\!\rangle}{\langle\!\langle {\rho} | {\rho} \rangle\!\rangle} .
\end{equation}
Figure~\ref{fig:S2}(a) shows the result.
In the initial Kitaev bilayer product state ($t=0$), we have $\langle\!\langle \sigma_j^\alpha \sigma_k^\alpha \tau_j^\alpha \tau_k^\alpha \rangle\!\rangle=\langle\!\langle \sigma_j^\alpha \sigma_k^\alpha \rangle\!\rangle \langle\!\langle \tau_j^\alpha \tau_k^\alpha \rangle\!\rangle
$. This quantity gradually increases and converges to the expected value $\langle\!\langle \sigma_j^\alpha \sigma_k^\alpha \tau_j^\alpha \tau_k^\alpha \rangle\!\rangle=1$, i.e., $\langle\!\langle \phi_j^\alpha \phi_k^\alpha \rangle\!\rangle=1$.
This confirms that the steady state has the property $\phi_j^\alpha=\phi_k^\alpha(=\pm 1)$ at each bond $\langle jk \rangle_\alpha$, allowing us to define the $\mathbb{Z}_2$ variable $X_{jk} (\equiv \phi_j^\alpha=\phi_k^\alpha)$ to characterize the steady state.

We also calculated the entanglement entropy between the two layers,
\begin{equation}
S_{\rm layer}=-\log ({\rm Tr}_{\{\sigma\}} \rho_{\rm layer}^2),
\end{equation}
where $\rho_{\rm layer}={\rm Tr}_{\{\tau\}}|\tilde{\rho}\rangle\!\rangle \langle\!\langle \tilde{\rho} |$ is the reduced density matrix obtained by tracing out the $\tau$-spins in the normalized wave function $|\tilde{\rho}\rangle\!\rangle= |\rho\rangle\!\rangle/\sqrt{\langle\!\langle \rho | \rho \rangle\!\rangle}$.
Figure~\ref{fig:S2}(b) shows the calculated entanglement entropy.
Starting from zero in the initial Kitaev bilayer product state, the entanglement entropy reaches the maximum value $S_{\rm layer}=2^{N/2-1}$ in the steady state. The steady state entanglement entropy is in agreement with the number of all possible states satisfying the steady state condition $\phi_j^\alpha \phi_k^\alpha=1$ [Eq.~(\ref{eq:steady-state-condition})].

\section{Toric code open quantum systems\label{appendix:toric_code}}

The open quantum system of the toric code model can be exactly solved in the steady state limit.
We find the exact steady state solution by taking the mapping to the non-Hermitian bilayer system,
\begin{eqnarray}
\mathpzc{H}_{\rm TC}
&=&
-J_A \sum_s A_s^{(\sigma)} - J_B \sum_p B_p^{(\sigma)} 
\nonumber\\
&&
+J_A \sum_s A_s^{(\tau)} + J_B \sum_p B_p^{(\tau)}
\nonumber\\
&&
+i \gamma \sum_j (\sigma_j^x \tau_j^x - 1),
\end{eqnarray}
where the upper layer and lower layer terms are denoted by the superscripts $(\sigma,\tau)$, respectively, and
\begin{eqnarray}
A_s^{(\sigma)} = \prod_{j\in s} \sigma_j^x,
~~~
B_p^{(\sigma)} = \prod_{j\in p} \sigma_j^z,
\\
A_s^{(\tau)} = \prod_{j\in s} \tau_j^x,
~~~
B_p^{(\tau)} = \prod_{j\in p} \tau_j^z.
\end{eqnarray}
We are going to use the basis shown in Table~\ref{tab:SI}.
In this basis, spin operators are given by
\begin{eqnarray}
\sigma^x
&=& 
 | s \rangle \langle t_x | 
+ | t_x \rangle \langle s | 
-i | t_y \rangle \langle t_z | 
+i | t_z \rangle \langle t_y | ,
\\
\tau^x
&=&
- | s \rangle \langle t_x | 
- | t_x \rangle \langle s | 
-i | t_y \rangle \langle t_z | 
+i | t_z \rangle \langle t_y | .
\end{eqnarray}
Other spin operators are obtained by cyclic permutations of the above.

\begin{table}[tb]
\centering
\begin{tabular}{cc}
\hline
$\sigma^z \otimes \tau^z$ basis & $|\phi^x,\phi^y,\phi^z\rangle$ basis
\\
\hline
$| s \rangle=\frac{1}{\sqrt{2}} ( |\uparrow \downarrow \rangle - | \downarrow \uparrow \rangle )$ & $|-1,-1,-1\rangle$
\\
$| t_x \rangle=\frac{-1}{\sqrt{2}} ( |\uparrow \uparrow \rangle - | \downarrow \downarrow \rangle )$ & $|-1,+1,+1\rangle$
\\
$| t_y \rangle=\frac{i}{\sqrt{2}} ( |\uparrow \uparrow \rangle + | \downarrow \downarrow \rangle )$ & $|+1,-1,+1\rangle$
\\
$| t_z \rangle=\frac{1}{\sqrt{2}} ( |\uparrow \downarrow \rangle + | \downarrow \uparrow \rangle )$ & $|+1,+1,-1\rangle$
\\
\hline
\end{tabular}
\caption{Four local states of  $\sigma$ and $\tau$ spins.
The four states $\{ |s\rangle, |t_x\rangle, |t_y\rangle, |t_z\rangle \}$ are eigenstates of $\phi^{x,y,z}$ (where $\phi^{\alpha}=\sigma^{\alpha}\tau^{\alpha}$).}
\label{tab:SI}
\end{table}

The steady state solution of the non-Hermitian Hamiltonian $\mathpzc{H}_{\rm TC}$ takes the following form:
\begin{equation}
| \rho_{\rm ss} \rangle\!\rangle = \mathcal{N} 
\left( \prod_s \frac{1+A_s^{(\sigma)}}{2} \right) \otimes_l | t_y \rangle_l ,
\end{equation}
where $\mathcal{N}$ is a normalization constant.
Now we show that $| \rho_{\rm ss} \rangle\!\rangle$ is indeed the steady state ($\mathpzc{H}_{\rm TC}| \rho_{\rm ss} \rangle\!\rangle=0$).
First, we note the property
\begin{equation}
\sigma_j^x \tau_j^x | \rho_{\rm ss} \rangle\!\rangle = | \rho_{\rm ss} \rangle\!\rangle ,
\label{eq:property1}
\end{equation}
which is easily checked from the identity $ \sigma^x \tau^x | t_y \rangle = \phi^x | t_y \rangle = | t_y \rangle$.
From the above property, we find that
\begin{equation}
A_s^{(\sigma)} | \rho_{\rm ss} \rangle\!\rangle = A_s^{(\tau)}  | \rho_{\rm ss} \rangle\!\rangle = | \rho_{\rm ss} \rangle\!\rangle .
\label{eq:property2}
\end{equation}
Lastly, we note that
\begin{equation}
B_p^{(\sigma)} | \rho_{\rm ss} \rangle\!\rangle = B_p^{(\tau)} | \rho_{\rm ss} \rangle\!\rangle ,
\label{eq:property3}
\end{equation}
which is derived from the identity $ \sigma^z | t_y \rangle = -i | t_x \rangle = \tau^z | t_y \rangle$.
The three properties of Eqs.~(\ref{eq:property1}),(\ref{eq:property2}),(\ref{eq:property3}) prove that $\mathpzc{H}_{\rm TC}| \rho_{\rm ss} \rangle\!\rangle=0$.

We go back to the original system by applying the inverse vectorization $|t_y\rangle \Rightarrow i \mathbb{I}_2/\sqrt{2}$ ($\mathbb{I}_2$: 2$\times$2 identity matrix) to the state $| \rho_{\rm ss} \rangle\!\rangle$.
Then, we obtain the steady state density matrix shown in Eq.~(\ref{eq:TC-steady-state}):
\begin{equation}
\hat{\rho}_{\rm ss} = \mathcal{N} \left( \frac{i}{\sqrt{2}} \right)^{N} 
\left( \prod_s \frac{1+A_s}{2} \right) ,
\end{equation}
where $N$ is the number of sites.
This is the projection operator into the subspace of $\{A_s=1\}$, leading to another representation
\begin{equation}
\hat{\rho}_{\rm ss} \propto \sum_n | \Psi_n^{(m)} \rangle \langle \Psi_n^{(m)} | ,
\end{equation}
i.e., the maximally mixed state of all possible $m$-particle excitations (with no $e$-particles) with equal weight.
This implies that the state $| \rho_{\rm ss} \rangle\!\rangle \propto \sum_n | \Psi_n^{(m)} \rangle \otimes| \Psi_n^{(m)} \rangle$ is an $m$-particle-condensed state of the $\mathbb{Z}_2\times\mathbb{Z}_2$ toric code bilayer.

On the other hand, the state $| \rho_{\rm ss} \rangle\!\rangle$ can be represented in the form of a $\mathbb{Z}_2$ toric code state.
To show this, we define a new set of Pauli matrices in the space of $|t_{y}\rangle$ and $|t_{z}\rangle$.
\begin{equation}
X \equiv | t_y \rangle \langle \tilde{t}_z | + | \tilde{t}_z \rangle \langle t_y |
~~~\&~~~
Z \equiv | t_y \rangle \langle t_y | - | \tilde{t}_z \rangle \langle \tilde{t}_z | ,
\end{equation}
where we used $| \tilde{t}_z \rangle = i | t_z \rangle$ instead of $| t_z \rangle$.
Using the new operators, we can rewrite the state as
\begin{equation}
| \rho_{\rm ss} \rangle\!\rangle = \mathcal{N} 
\left( \prod_s \frac{1+A_s^{(X)}}{2} \right) \otimes_l | Z=1 \rangle_l .
\end{equation}
We may also write it as
\begin{equation}
| \rho_{\rm ss} \rangle\!\rangle = \mathcal{N} 
\left( \prod_p \frac{1+B_p^{(Z)}}{2} \right) \otimes_l | X=1 \rangle_l .
\end{equation}
This is exactly the $\mathbb{Z}_2$ toric code state of $\{A_s=1~\&~B_p=1\}$.
All these results confirm the anyon condensation transition summarized in Eq.~(\ref{eq:TC-DAC}).

\bibliographystyle{quantum}

\begin{thebibliography}{10}

\bibitem{Kitaev2006}
A.~Kitaev.
\newblock ``{Anyons in an exactly solved model and beyond}''.
\newblock
  \href{https://dx.doi.org/https://doi.org/10.1016/j.aop.2005.10.005}{Ann.
  Phys. {\bf 321}, 2--111}~(2006).

\bibitem{Kitaev2003}
A.Yu. Kitaev.
\newblock ``{Fault-tolerant quantum computation by anyons}''.
\newblock
  \href{https://dx.doi.org/https://doi.org/10.1016/S0003-4916(02)00018-0}{Ann.
  Phys. {\bf 303}, 2--30}~(2003).

\bibitem{WenBook}
X.-G. Wen.
\newblock ``{Quantum Field Theory of Many-Body Systems: From the Origin of
  Sound to an Origin of Light and Electrons}''.
\newblock
  \href{https://dx.doi.org/10.1093/acprof:oso/9780199227259.001.0001}{Oxford
  University Press}. ~(2007).

\bibitem{SachdevBook}
S.~Sachdev.
\newblock ``{Quantum Phases of Matter}''.
\newblock
  \href{https://dx.doi.org/https://doi.org/10.1017/9781009212717}{Cambridge
  University Press}. ~(2023).

\bibitem{Savary2016}
L.~Savary and L.~Balents.
\newblock ``{Quantum spin liquids: a review}''.
\newblock \href{https://dx.doi.org/10.1088/0034-4885/80/1/016502}{Rep. Prog.
  Phys. {\bf 80}, 016502}~(2016).

\bibitem{Broholm2020}
C.~Broholm, R.~J. Cava, S.~A. Kivelson, D.~G. Nocera, M.~R. Norman, and
  T.~Senthil.
\newblock ``{Quantum spin liquids}''.
\newblock \href{https://dx.doi.org/10.1126/science.aay0668}{Science {\bf 367},
  eaay0668}~(2020).

\bibitem{Knolle2019}
J.~Knolle and R.~Moessner.
\newblock ``{A Field Guide to Spin Liquids}''.
\newblock
  \href{https://dx.doi.org/10.1146/annurev-conmatphys-031218-013401}{Annu. Rev.
  Condens. Matter Phys. {\bf 10}, 451--472}~(2019).

\bibitem{Kasahara2018}
Y.~Kasahara, T.~Ohnishi, Y.~Mizukami, O.~Tanaka, S.~Ma, K.~Sugii, N.~Kurita,
  H.~Tanaka, J.~Nasu, Y.~Motome, et~al.
\newblock ``{Majorana quantization and half-integer thermal quantum Hall effect
  in a Kitaev spin liquid}''.
\newblock \href{https://dx.doi.org/10.1038/s41586-018-0274-0}{Nature {\bf 559},
  227--231}~(2018).

\bibitem{Takagi2019}
H.~Takagi, T.~Takayama, G.~Jackeli, G.~Khaliullin, and S.~E Nagler.
\newblock ``{Concept and realization of Kitaev quantum spin liquids}''.
\newblock \href{https://dx.doi.org/10.1038/s42254-019-0038-2}{Nat. Rev. Phys.
  {\bf 1}, 264--280}~(2019).

\bibitem{Motome2019}
Y.~Motome and J.~Nasu.
\newblock ``{Hunting Majorana Fermions in Kitaev Magnets}''.
\newblock \href{https://dx.doi.org/10.7566/JPSJ.89.012002}{J. Phys. Soc. Jpn.
  {\bf 89}, 012002}~(2019).

\bibitem{Trebst2022}
S.~Trebst and C.~Hickey.
\newblock ``{Kitaev materials}''.
\newblock
  \href{https://dx.doi.org/https://doi.org/10.1016/j.physrep.2021.11.003}{Physics
  Reports {\bf 950}, 1--37}~(2022).

\bibitem{Preskill2002}
E.~Dennis, A.~Kitaev, A.~Landahl, and J.~Preskill.
\newblock ``{Topological quantum memory}''.
\newblock \href{https://dx.doi.org/10.1063/1.1499754}{Journal of Mathematical
  Physics {\bf 43}, 4452--4505}~(2002).

\bibitem{Bombin2008}
H.~Bombin and M.~A. Martin-Delgado.
\newblock ``{Family of non-Abelian Kitaev models on a lattice: Topological
  condensation and confinement}''.
\newblock \href{https://dx.doi.org/10.1103/PhysRevB.78.115421}{Phys. Rev. B
  {\bf 78}, 115421}~(2008).

\bibitem{Bombin2010}
H.~Bombin.
\newblock ``{Topological Order with a Twist: Ising Anyons from an Abelian
  Model}''.
\newblock \href{https://dx.doi.org/10.1103/PhysRevLett.105.030403}{Phys. Rev.
  Lett. {\bf 105}, 030403}~(2010).

\bibitem{Pachosbook}
J.~K. Pachos.
\newblock ``{Introduction to Topological Quantum Computation}''.
\newblock \href{https://dx.doi.org/DOI: 10.1017/CBO9780511792908}{Cambridge
  University Press}. Cambridge~(2012).

\bibitem{Terhal2015}
B.~M. Terhal.
\newblock ``{Quantum error correction for quantum memories}''.
\newblock \href{https://dx.doi.org/10.1103/RevModPhys.87.307}{Rev. Mod. Phys.
  {\bf 87}, 307--346}~(2015).

\bibitem{Jian-WeiPan2017}
H.-N. Dai, B.~Yang, A.~Reingruber, H.~Sun, X.-F. Xu, Y.-A. Chen, Z.-S. Yuan,
  and J.-W. Pan.
\newblock ``{Four-body ring-exchange interactions and anyonic statistics within
  a minimal toric-code Hamiltonian}''.
\newblock \href{https://dx.doi.org/10.1038/nphys4243}{Nat. Phys. {\bf 13},
  1195--1200}~(2017).

\bibitem{Satzinger2021}
K.~J. Satzinger, Y.-J. Liu, A.~Smith, C.~Knapp, M.~Newman, C.~Jones, Z.~Chen,
  C.~Quintana, X.~Mi, A.~Dunsworth, et~al.
\newblock ``{Realizing topologically ordered states on a quantum processor}''.
\newblock \href{https://dx.doi.org/10.1126/science.abi8378}{Science {\bf 374},
  1237--1241}~(2021).

\bibitem{Semeghini2021}
G.~Semeghini, H.~Levine, A.~Keesling, S.~Ebadi, T.~T. Wang, D.~Bluvstein,
  R.~Verresen, H.~Pichler, M.~Kalinowski, R.~Samajdar, et~al.
\newblock ``{Probing topological spin liquids on a programmable quantum
  simulator}''.
\newblock \href{https://dx.doi.org/10.1126/science.abi8794}{Science {\bf 374},
  1242--1247}~(2021).

\bibitem{Samajdar2020}
R.~Samajdar, W.~W. Ho, H.~Pichler, M.~D. Lukin, and S.~Sachdev.
\newblock ``{Quantum phases of Rydberg atoms on a kagome lattice}''.
\newblock \href{https://dx.doi.org/10.1073/pnas.2015785118}{Proc. Natl. Acad.
  Sci. U.S.A. {\bf 118}, e2015785118}~(2021).

\bibitem{Verresen2021}
R.~Verresen, M.~D. Lukin, and A.~Vishwanath.
\newblock ``{Prediction of Toric Code Topological Order from Rydberg
  Blockade}''.
\newblock \href{https://dx.doi.org/10.1103/PhysRevX.11.031005}{Phys. Rev. X
  {\bf 11}, 031005}~(2021).

\bibitem{Williamson2022}
T.~D. Ellison, Y.-A. Chen, A.~Dua, W.~Shirley, N.~Tantivasadakarn, and D.~J.
  Williamson.
\newblock ``{Pauli Stabilizer Models of Twisted Quantum Doubles}''.
\newblock \href{https://dx.doi.org/10.1103/PRXQuantum.3.010353}{PRX Quantum
  {\bf 3}, 010353}~(2022).

\bibitem{Pollmann2022}
Y.-J. Liu, K.~Shtengel, A.~Smith, and F.~Pollmann.
\newblock ``{Methods for Simulating String-Net States and Anyons on a Digital
  Quantum Computer}''.
\newblock \href{https://dx.doi.org/10.1103/PRXQuantum.3.040315}{PRX Quantum
  {\bf 3}, 040315}~(2022).

\bibitem{Hsieh2022}
T.-C. Lu, L.~A. Lessa, I.~H. Kim, and T.~H. Hsieh.
\newblock ``{Measurement as a Shortcut to Long-Range Entangled Quantum
  Matter}''.
\newblock \href{https://dx.doi.org/10.1103/PRXQuantum.3.040337}{PRX Quantum
  {\bf 3}, 040337}~(2022).

\bibitem{Tantivasadakarn2023}
N.~Tantivasadakarn, A.~Vishwanath, and R.~Verresen.
\newblock ``{Hierarchy of Topological Order From Finite-Depth Unitaries,
  Measurement, and Feedforward}''.
\newblock \href{https://dx.doi.org/10.1103/PRXQuantum.4.020339}{PRX Quantum
  {\bf 4}, 020339}~(2023).

\bibitem{Google2023}
Google~Quantum AI and Collaborators.
\newblock ``{Non-Abelian braiding of graph vertices in a superconducting
  processor}''.
\newblock \href{https://dx.doi.org/10.1038/s41586-023-05954-4}{Nature {\bf
  618}, 264--269}~(2023).

\bibitem{Verresen2022}
R.~Verresen, N.~Tantivasadakarn, and A.~Vishwanath.
\newblock ``{Efficiently preparing Schr\"odinger's cat, fractons and
  non-Abelian topological order in quantum devices}''~(2022).
\newblock  \href{http://arxiv.org/abs/2112.03061}{arXiv:2112.03061}.

\bibitem{Bravyi2022}
S.~Bravyi, I.~Kim, A.~Kliesch, and R.~Koenig.
\newblock ``{Adaptive constant-depth circuits for manipulating non-abelian
  anyons}''~(2022).
\newblock  \href{http://arxiv.org/abs/2205.01933}{arXiv:2205.01933}.

\bibitem{Tantivasadakarn2022shortest}
N.~Tantivasadakarn, R.~Verresen, and A.~Vishwanath.
\newblock ``The shortest route to non-abelian topological order on a quantum
  processor''.
\newblock  \href{https://doi.org/10.1103/PhysRevLett.131.060405}{Phys. Rev. Lett. 131, 060405}~(2023).

\bibitem{Kim2023}
M.~Foss-Feig, A.~Tikku, T.-C. Lu, K.~Mayer, M.~Iqbal, T.~M. Gatterman, J.~A.
  Gerber, K.~Gilmore, D.~Gresh, A.~Hankin, N.~Hewitt, C.~V. Horst, M.~Matheny,
  T.~Mengle, B.~Neyenhuis, H.~Dreyer, D.~Hayes, T.~H. Hsieh, and I.~H. Kim.
\newblock ``{Experimental demonstration of the advantage of adaptive quantum
  circuits}''~(2023).
\newblock  \href{http://arxiv.org/abs/2302.03029}{arXiv:2302.03029}.

\bibitem{Iqbal2023}
M.~Iqbal, N.~Tantivasadakarn, R.~Verresen, S.~L. Campbell, J.~M. Dreiling,
  C.~Figgatt, J.~P. Gaebler, J.~Johansen, M.~Mills, S.~A. Moses, J.~M. Pino,
  A.~Ransford, M.~Rowe, P.~Siegfried, R.~P. Stutz, M.~Foss-Feig, A.~Vishwanath,
  and H.~Dreyer.
\newblock ``{Non-Abelian topological order and anyons on a trapped-ion processor}''.
\newblock  \href{https://doi.org/10.1038/s41586-023-06934-4}{Nature {\bf 626}, 505–511}~(2024). 

\bibitem{Preskill2018}
J.~Preskill.
\newblock ``Quantum {C}omputing in the {NISQ} era and beyond''.
\newblock \href{https://dx.doi.org/10.22331/q-2018-08-06-79}{{Quantum} {\bf 2},
  79}~(2018).

\bibitem{Zoller2008}
B.~Kraus, H.~P. B\"uchler, S.~Diehl, A.~Kantian, A.~Micheli, and P.~Zoller.
\newblock ``{Preparation of entangled states by quantum Markov processes}''.
\newblock \href{https://dx.doi.org/10.1103/PhysRevA.78.042307}{Phys. Rev. A
  {\bf 78}, 042307}~(2008).

\bibitem{Diehl2008natphys}
S.~Diehl, A.~Micheli, A.~Kantian, B.~Kraus, H.~P. B{\"u}chler, and P.~Zoller.
\newblock ``{Quantum states and phases in driven open quantum systems with cold
  atoms}''.
\newblock \href{https://dx.doi.org/10.1038/nphys1073}{Nat. Phys. {\bf 4},
  878--883}~(2008).

\bibitem{Diehl2011natphys}
S.~Diehl, E.~Rico, M.~A. Baranov, and P.~Zoller.
\newblock ``{Topology by dissipation in atomic quantum wires}''.
\newblock \href{https://dx.doi.org/10.1038/nphys2106}{Nat. Phys. {\bf 7},
  971--977}~(2011).

\bibitem{Lindblad1976}
G.~Lindblad.
\newblock ``{On the generators of quantum dynamical semigroups}''.
\newblock \href{https://dx.doi.org/10.1007/BF01608499}{Communications in
  Mathematical Physics {\bf 48}, 119--130}~(1976).

\bibitem{Gorini1976}
V.~Gorini, A.~Kossakowski, and E.~C.~G. Sudarshan.
\newblock ``{Completely positive dynamical semigroups of N-level systems}''.
\newblock \href{https://dx.doi.org/10.1063/1.522979}{Journal of Mathematical
  Physics {\bf 17}, 821--825}~(1976).

\bibitem{BreuerPetruccione}
H.-P. Breuer and F.~Petruccione.
\newblock ``{{The Theory of Open Quantum Systems}}''.
\newblock
  \href{https://dx.doi.org/10.1093/acprof:oso/9780199213900.001.0001}{Oxford
  University Press}. ~(2007).

\bibitem{Albert2014}
V.~V. Albert and L.~Jiang.
\newblock ``{Symmetries and conserved quantities in Lindblad master
  equations}''.
\newblock \href{https://dx.doi.org/10.1103/PhysRevA.89.022118}{Phys. Rev. A
  {\bf 89}, 022118}~(2014).

\bibitem{Albert2016}
V.~V. Albert, B.~Bradlyn, M.~Fraas, and L.~Jiang.
\newblock ``{Geometry and Response of Lindbladians}''.
\newblock \href{https://dx.doi.org/10.1103/PhysRevX.6.041031}{Phys. Rev. X {\bf
  6}, 041031}~(2016).

\bibitem{Lidar2020}
D.~A. Lidar.
\newblock ``{Lecture Notes on the Theory of Open Quantum Systems}''~(2020).
\newblock  \href{http://arxiv.org/abs/1902.00967}{arXiv:1902.00967}.

\bibitem{Harrington2022}
P.~M. Harrington, E.~J. Mueller, and K.~W. Murch.
\newblock ``{Engineered dissipation for quantum information science}''.
\newblock \href{https://dx.doi.org/10.1038/s42254-022-00494-8}{Nat. Rev. Phys.
  {\bf 4}, 660--671}~(2022).

\bibitem{Bais2009}
F.~A. Bais and J.~K. Slingerland.
\newblock ``{Condensate-induced transitions between topologically ordered
  phases}''.
\newblock \href{https://dx.doi.org/10.1103/PhysRevB.79.045316}{Phys. Rev. B
  {\bf 79}, 045316}~(2009).

\bibitem{Burnell2011}
F.~J. Burnell, S.~H. Simon, and J.~K. Slingerland.
\newblock ``{Condensation of achiral simple currents in topological lattice
  models: Hamiltonian study of topological symmetry breaking}''.
\newblock \href{https://dx.doi.org/10.1103/PhysRevB.84.125434}{Phys. Rev. B
  {\bf 84}, 125434}~(2011).

\bibitem{Burnell2018}
F.~J. Burnell.
\newblock ``{Anyon Condensation and Its Applications}''.
\newblock
  \href{https://dx.doi.org/10.1146/annurev-conmatphys-033117-054154}{Annu. Rev.
  Condens. Matter Phys. {\bf 9}, 307--327}~(2018).

\bibitem{Schmidt2020}
R.~Wiedmann, L.~Lenke, M.~R. Walther, M.~M\"uhlhauser, and K.~P. Schmidt.
\newblock ``{Quantum critical phase transition between two topologically
  ordered phases in the Ising toric code bilayer}''.
\newblock \href{https://dx.doi.org/10.1103/PhysRevB.102.214422}{Phys. Rev. B
  {\bf 102}, 214422}~(2020).

\bibitem{Hwang2024}
K.~Hwang.
\newblock ``{Anyon condensation and confinement transition in a Kitaev spin liquid bilayer}''~(2024).
\newblock  \href{https://doi.org/10.1103/PhysRevB.109.134412}{Phys. Rev. B
  {\bf 109}, 134412}~(2024).

\bibitem{Katsura2019}
N.~Shibata and H.~Katsura.
\newblock ``{Dissipative spin chain as a non-Hermitian Kitaev ladder}''.
\newblock \href{https://dx.doi.org/10.1103/PhysRevB.99.174303}{Phys. Rev. B
  {\bf 99}, 174303}~(2019).

\bibitem{Katsura2019-2}
N.~Shibata and H.~Katsura.
\newblock ``{Dissipative quantum Ising chain as a non-Hermitian Ashkin-Teller
  model}''.
\newblock \href{https://dx.doi.org/10.1103/PhysRevB.99.224432}{Phys. Rev. B
  {\bf 99}, 224432}~(2019).

\bibitem{Saad1992}
Y.~Saad.
\newblock ``{Analysis of Some Krylov Subspace Approximations to the Matrix
  Exponential Operator}''.
\newblock \href{https://dx.doi.org/10.1137/0729014}{SIAM Journal on Numerical
  Analysis {\bf 29}, 209--228}~(1992).

\bibitem{sidje1998expokit}
R.~B. Sidje.
\newblock ``{Expokit: A software package for computing matrix exponentials}''.
\newblock \href{https://dx.doi.org/https://doi.org/10.1145/285861.285868}{ACM
  Transactions on Mathematical Software (TOMS) {\bf 24}, 130--156}~(1998).

\bibitem{Jamiolkowski1972}
A.~Jamio{\l}kowski.
\newblock ``{Linear transformations which preserve trace and positive
  semidefiniteness of operators}''.
\newblock
  \href{https://dx.doi.org/https://doi.org/10.1016/0034-4877(72)90011-0}{Reports
  on Mathematical Physics {\bf 3}, 275--278}~(1972).

\bibitem{Choi1975}
M.-D. Choi.
\newblock ``{Completely positive linear maps on complex matrices}''.
\newblock
  \href{https://dx.doi.org/https://doi.org/10.1016/0024-3795(75)90075-0}{Linear
  Algebra and its Applications {\bf 10}, 285--290}~(1975).

\bibitem{Yang2021}
K.~Yang, S.~C.~Morampudi, and E.~J.~Bergholtz.
\newblock ``{Exceptional Spin Liquids from Couplings to the Environment}''.
\newblock 
\href{https://doi.org/10.1103/PhysRevLett.126.077201}{Phys. Rev. Lett.
 {\bf 126}, 077201}~(2021).
%
\bibitem{Lee2023}
J.~Y. Lee, C.-M. Jian, and C.~Xu.
\newblock ``{Quantum Criticality Under Decoherence or Weak Measurement}''.
\newblock  \href{https://doi.org/10.1103/PRXQuantum.4.030317}{PRX Quantum {\bf 4}, 030317}~(2023).
%
\bibitem{Bao2023}
Y.~Bao, R.~Fan, A.~Vishwanath, and E.~Altman.
\newblock ``{Mixed-state topological order and the errorfield double
  formulation of decoherence-induced transitions}''~(2023).
\newblock  \href{http://arxiv.org/abs/2301.05687}{arXiv:2301.05687}.
%
\bibitem{Fan2024}
R.~Fan, Y.~Bao, E.~Altman, and A.~Vishwanath.
\newblock ``{Diagnostics of Mixed-State Topological Order and Breakdown of Quantum Memory}''.
\newblock  \href{https://doi.org/10.1103/PRXQuantum.5.020343}{PRX Quantum {\bf 5}, 020343}~(2024).
%
\bibitem{Wang2023}
Z.~Wang, Z.~Wu, and Z.~Wang.
\newblock ``{Intrinsic Mixed-state Quantum Topological Order}''~(2023).
\newblock  \href{https://arxiv.org/abs/2307.13758}{arXiv:2307.13758}.
%
\bibitem{Sohal2024}
R.~Sohal and A.~Prem.
\newblock ``{A Noisy Approach to Intrinsically Mixed-State Topological Order
}''~(2024).
\newblock  \href{https://arxiv.org/abs/2403.13879}{arXiv:2403.13879}.
%
\bibitem{Ellison2024}
T.~Ellison and M.~Cheng.
\newblock ``{Towards a classification of mixed-state topological orders in two dimensions}''~(2024).
\newblock  \href{https://arxiv.org/abs/2405.02390}{arXiv:2405.02390}.


\end{thebibliography}

\end{document}